\documentclass[aps,prl,a4paper,reprint,showpacs,amsmath,amssymb,superscriptaddress,floatfix]{revtex4-1}

\usepackage{amssymb}
\usepackage{latexsym}
\usepackage[dvips]{graphicx}
\usepackage{epsfig}

\usepackage{dcolumn}
\usepackage{amsmath}
\usepackage{amsfonts}
\usepackage{bm}
\usepackage{color}

\newcommand{\be}{\begin{equation}}
\newcommand{\ee}{\end{equation}}
\newcommand{\bea}{\begin{eqnarray}}
\newcommand{\eea}{\end{eqnarray}}

\newcommand{\p}{\partial}

\newcommand{\la}{\langle}
\newcommand{\ra}{\rangle}

\begin{document}

\title{Kondo Impurities Coupled to Helical Luttinger Liquid: \\
        RKKY-Kondo Physics Revisited}

\author{Oleg M. Yevtushenko}
\affiliation{Ludwig Maximilians University, Arnold Sommerfeld Center and Center for Nano-Science,
                 Munich, DE-80333, Germany}

\author{Vladimir I. Yudson}
\affiliation{Laboratory for Condensed Matter Physics, National Research University Higher School of Economics, Moscow, 101000, Russia}
\affiliation{Russian Quantum Center, Skolkovo, Moscow Region 143025, Russia}

\date{\today }

\begin{abstract}
We show that the paradigmatic Ruderman--Kittel--Kasuya--Yosida (RKKY) description of
two local magnetic moments coupled to propagating electrons breaks down in helical
Luttinger Liquids when the electron interaction is stronger than some critical value.
In this novel regime, the Kondo effect overwhelms the RKKY interaction over all macroscopic
inter-impurity distances. This phenomenon is a direct consequence of the helicity (realized,
for instance, at edges of a time-reversal invariant topological insulator) and does not take
place in usual (non-helical) Luttinger Liquids.
\end{abstract}


\maketitle

The seminal problem of the indirect exchange interaction (RKKY) between two spatially
localized magnetic moments, i.e. Kondo impurities (KIs), weakly coupled to propagating
electrons has the well-known solution  \cite{Kittel}. The paradigmatic approach can be
reformulated in the contemporary language as follows: one integrates out fermionic degrees
of freedom and reduces the resulting non-local Lagrangian to the effective spin Hamiltonian.
The second step is usually justified by a scale separation, the spin dynamics is slower
than the electron one if the electron-spin coupling is weak.
RKKY induces perceptible inter-impurity correlations if an inter-impurity distance,
$ R $, is smaller then the thermal length and the electron coherence length.
This RKKY theory is the obvious simplification since it neglects
another fundamental phenomenon, namely,  the Kondo effect \cite{Hewson}. If the temperature
is below the Kondo temperature, $ T < T_{K} $, the antiferromagnetic Kondo coupling drives the
single KI to the strong coupling limit where the electrons screen KI. Hence, the Kondo screening is
an antagonist of the RKKY interaction.

The RKKY-Kondo interplay has attracted a large attention since several decades
\cite{jones_TwoImpKondo_1988,fye_TwoImpKondo_1989,fye_TwoImpKondo_1994,gan_TwoImpKondo_1995,affleck_TwoImpKondo_1995}
and remains a hot topic of research because of its importance for new systems,
such as graphene \cite{Kogan_2011,allerdt_graphene_2017}, strongly correlated quantum wires and carbon nanotubes,
which are described by the Luttinger Liquid model \cite{RKKY-LL1,RKKY-LL2}. The latter are especially
interesting because
the Kondo effect can be enhanced by the interactions \cite{LeeToner,FuruNaga,EggerKomnik}. The
common wisdom is that the RKKY physics dominates in a broad macroscopic range of $ R $ 
in three-  and low-dimensional systems.

{\it In this Letter}, we will demonstrate that, surprisingly, the paradigmatic RKKY approach 
breaks down in strongly correlated helical systems - Helical Luttinger Liquids (HLLs). We 
will show that the reason of this unexpected finding is the nontrivial and unusually
increased RKKY-Kondo competition.

Helicity means the lock-in relation between electron spin and momentum: helical
electrons propagating in opposite directions have opposite spins. This protects the helical
transport against effects of spinless impurities.
HLL can appear at edges of time-reversal invariant 2D topological insulators \cite{HasanKane,QiZhang,TI-Shen,WuBernevigZhang,XuMoore}
and in purely 1D interacting systems \cite{TsvYev2015,Schimmel2016}.
The Kondo effect \cite{MaciejkoOregZhang,FurusakiMatveev,MaciejkoLattice,Johannesson_2012,Feiguin_Martins_2017} and
RKKY \cite{RKKY-HLL,RKKY-3DTI-1,RKKY-3DTI-2,RKKY-3DTI-3,RKKY-HLL-Int,RKKY-HLL-Bulk} in
the topological insulators are intensively studied since past several years. This
increasing interest is partly related to the hypothesis that Kondo/RKKY effects can
be responsible for deviations of the helical conductance from its ideal value,
see Refs.\cite{CheiGlaz,AAY,Yevt-Helical,vayrynen_2016,Klinovaja_Loss_2017} and discussions therein.

At a simple phenomenological level, one can find ``the winner of the RKKY-Kondo competition''
by comparing $ T_{K} $ with the characteristic energy of RKKY, $ E_{\rm RKKY} $. The
latter has the meaning of the energy gap which opens after the RKKY correlations lift
a degeneracy in the energy of the uncorrelated KIs. In the absence of Coulomb interactions,
$ T^{(0)}_K \propto \exp(-1/\rho_0 J)$ and $ E^{(0)}_{\rm RKKY} \propto J^2 /R^d $;
where $ \rho_0 $ is the density of states of the electrons at the Fermi surface, $ J $
is the Kondo coupling constant and $ d $ is the space dimension. If $ \rho_0 J \ll 1 $,
there is a broad range of macroscopic distances where $ E^{(0)}_{\rm RKKY} \gg T^{(0)}_K, T $
and RKKY is expected to overwhelm the Kondo screening.

The situation drastically changes in HLL with the strong interaction. Let us concentrate on the XXZ
Kondo coupling with small constants $ J_\perp, J_z \ll 1/\rho_{0} $,
see the formal definition in Eqs.(\ref{L-fs},\ref{L-bs}), and temporarily neglect $ J_z $. The electron
repulsion is reflected by the Luttinger parameter of HLL: $ K \le 1 $ \cite{K-range}; $ K = 1 $ corresponds to
noninteracting fermions. Both, $ E_{\rm RKKY} $ [see Eq.(\ref{E_RKKY})] and $ T_{K} $ (see
Ref.\cite{MaciejkoOregZhang}), are modified by the interaction:
\bea
\label{Scale_RKKY}
   \quad E_{\rm RKKY}  & \sim & D \, (\rho_0 J_\perp)^2 \, (\xi/ R)^{2K-1}, \ 1/2 < K \le 1; \\
\label{Scale_Tk}
   T_{K} & \propto &
     \left\{
       \begin{array}{l}
          T_{K}^{(0)}, \ 0 < 1 - K \ll 1; \\
          D \, (\rho_0 J_\perp)^{\frac{1}{1-K}} \gg T_{K}^{(0)} \!\! , \ 1 - K \gg \rho_0 J_\perp .
       \end{array}
     \right.
\eea
Here $ \xi $ ($ D $) is the spatial (energy) UV cutoff, i.e. the lattice spacing (bandwidth).
A naive formal extension of Eq.(\ref{Scale_RKKY}) to the regime $ K < 1/2 $ would lead to a paradoxical
result: $ E_{\rm RKKY} $ seems to grow without bound with the increase of the inter-impurity distance.
The results presented below ultimately refute any possibility of such an effect.

\begin{figure}[t]
\begin{center}
   \includegraphics[width=0.48 \textwidth]{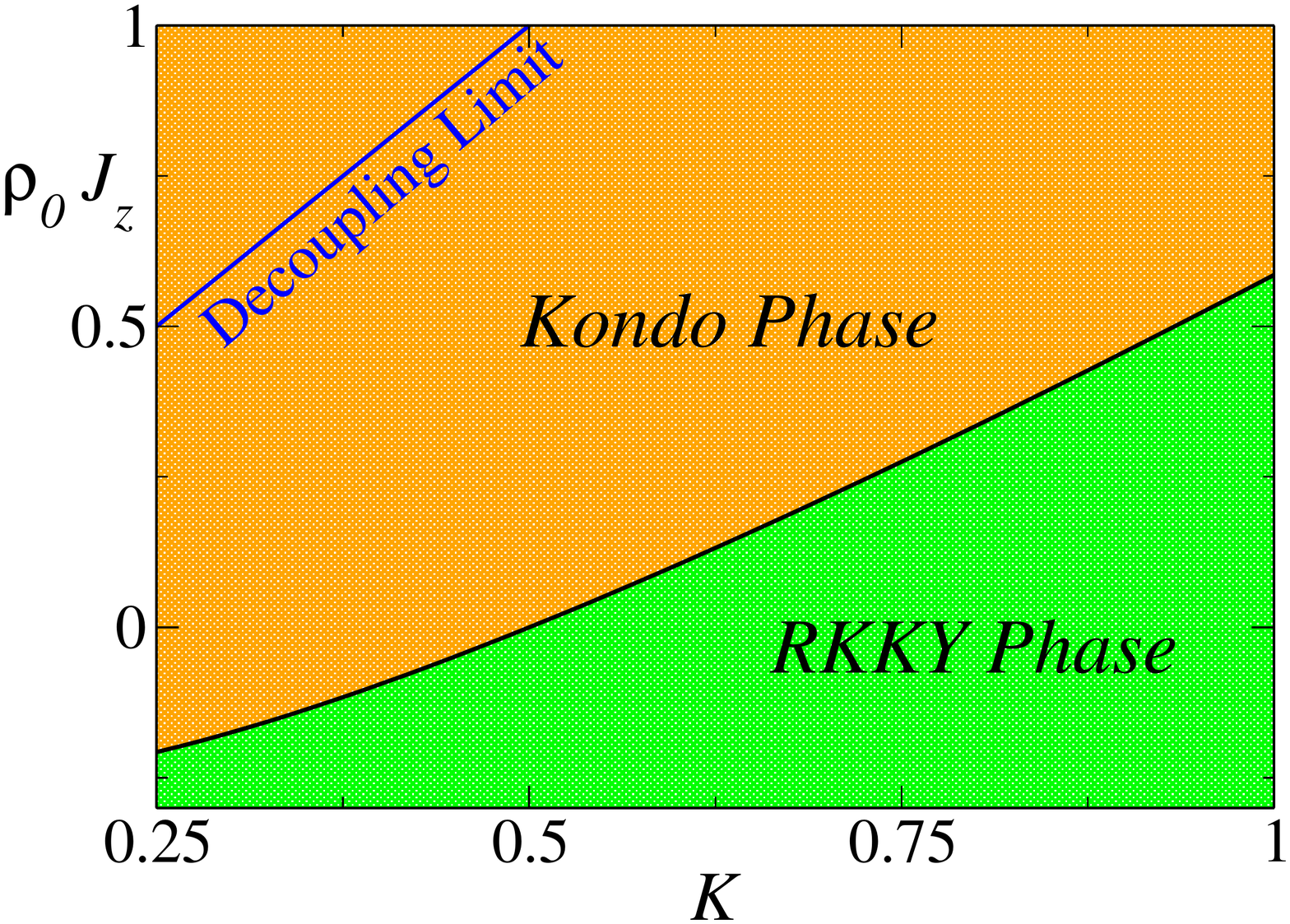}
\end{center}
\vspace{-0.5 cm}
   \caption{
        \label{PhaseDiagr}
        (color on-line) Phase diagram of the two Kondo impurities coupled to the HLL.
        Green and orange regions demonstrate the RKKY- and the Kondo- phases,
        respectively. Axes show values of the Luttinger parameter $ 1/4 \le K \le 1 $ and the
        dimensionless coupling constant $ |\rho_0 J_z| < 1 $, see Eqs.(\ref{HLL-Lagr},\ref{L-fs}).
        The critical line, which separates the phases, is defined by the equation $ \tilde{K}
        = 1/2 $. The decoupling limit corresponds to $ \tilde{K} = 0 $ \cite{MaciejkoLattice}.
           }
\end{figure}

Based on the above explained phenomenological arguments, we expect that, if $ E_{\rm RKKY}(R \sim \xi) > T_K $,
there exists a broad range of macroscopic distances where RKKY dominates over the Kondo effect. In the
opposite case, $ E_{\rm RKKY}(R \sim \xi) < T_K $, the Kondo physics dominates everywhere. The border between
these two phases is defined by the condition $ E_{\rm RKKY} \sim T_{K} $. We will show that it corresponds to the
critical value of the effective interaction parameter
\be
\label{K_Eff}
  \tilde{K} = K (1-\rho_0 J_z/2 K)^2, \ \tilde{K}_{\rm crit} = 1/2,
\ee
see the phase diagram in Fig.\ref{PhaseDiagr}  \cite{comm-K}. The paradigmatic RKKY theory is valid only at
$ \tilde{K} > 1/2 $ and fails at  $ \tilde{K} < 1/2 $.
Namely, the spin subsystem cannot be described by an effective
(RKKY-like) Hamiltonian at $ \tilde{K} < 1/2 $.
These statements, which are proven below at a more formal level,
are our {\it main result}.

The rest of this paper is organized as follows: Firstly, we will rederive the RKKY Hamiltonian by integrating out
HLL degrees of freedom and discuss the difference between helical and usual (not helical but spinfull)
cases. We will combine the microscopic diagrammatic
approach with one-loop Renormalization Group (RG) arguments to explain how RKKY stops Kondo
renormalizations and why the paradigmatic theory of RKKY is valid in the range $ 1/2 < \tilde{K} $
and fails at $ \tilde{K} < 1/2 $. By exploiting the extreme situation close to the decoupling limit \cite{MaciejkoLattice}
we will demonstrate that the strong effective interaction makes the RKKY-induced spin correlations irrelevant. We
thus could conclude that the physics is fully dominated by the Kondo effect at $ \tilde{K} < 1/2 $.

{\it The model}: we use functional integrals in the Matsubara formulation with the imaginary time $ \tau $.
The bosonized Lagrangian density of HLL \cite{Giamarchi,WuBernevigZhang,MaciejkoOregZhang,MaciejkoLattice,Yevt-Helical} is
\be
\label{HLL-Lagr}
   {\cal L}_{\rm HLL} = \bigl[ (\partial_\tau \phi)^2 + (u \partial_x \phi)^2 \bigr] / ( 2 \pi u K );
\ee
here $ u $ is the velocity of bosonic excitations. The electron-KI interaction is described by Lagrangians
of the forward/backward scattering:
\bea
\label{L-fs}
   {\cal L}_{\rm fs} & = & i J_z a_{\rm fs}/(\pi u K)\sum_{j=1,2} \delta(x-x_j) S^z_j \partial_\tau \phi; \\
\label{L-bs}
   {\cal L}_{\rm bs} & = & J_\perp /(2 \pi \xi)\sum_{j=1,2}\delta(x-x_j) \bigl[ S^+_j e^{-2 i a_{\rm bs} \phi} + c.c. \bigr] .
\eea
Here $ x_j $ are impurity positions with $ R = | x_1 - x_2 | $, $ S^\mu_j $ are fields
describing KI spin degrees of freedom and we have introduced auxiliary dimensionless constants
$ a_{\rm fs,bs} $ which are explained below.
Eqs.(\ref{HLL-Lagr}-\ref{L-bs}) describe the low energy physics, i.e. all fields are smooth on the scale
of $ \xi $. In particular, $ 2 k_F $-oscillations ($ k_F $ is the Fermi momentum) are eliminated
from $  {\cal L}_{\rm bs} $ by the spin rotation $ S_j^\pm e^{\mp 2 i k_F x_j} \to S_j^\pm $.
We note in passing that, unlike previously studied examples \cite{withoff_1990,ingersent_1996,lee_2012},
features of a single particle density of states and the precise level of the chemical potential are
unimportant for the RKKY/Kondo physics which we explore. This is the peculiarity of the interacting
1D systems described by the bosonization approach \cite{Giamarchi}.
We emphasize that the helicity of our model implies that it has only U(1) spin symmetry but
no SU(2) symmetry. We restrict ourselves to the case of spin-1/2 KIs and choose a parametrization for $ S $-fields
in terms of Grassmann fields corresponding to Dirac fermions \cite{schad_2015,schad_2016}:
\be
\label{Diracs}
  S_j^+ = (\bar{d}_j + d_j) \bar{c}_j, \
  S_j^z = \bar{c}_j c_j - 1/2.
\ee
Each Grassmann field has the usual dynamical Lagrangian $ {\cal L}_f = \bar{\psi}_j \p_\tau \psi_j,
\ \psi_j = \{ c_j, d_j \} $ \cite{Sz-const,Spins-RKKY}.

In the initial formulation, one chooses $ a_{\rm fs,bs} = 1 $, however, the gauge transformation of $ c $-fermions,
$ c_j \to c_j \exp(i \lambda \phi(x_j)) $, which is equivalent to the Emery-Ki\-vel\-son rotation
\cite{EmKivRes,MaciejkoLattice},  allows one to represent the theory in two extreme forms:
\bea
\label{Model-1}
   \mbox{Representation 1:} & \quad & a_{\rm fs} = 0, \ a_{bs} = 1 - \kappa; \\
\label{Model-2}
   \mbox{Representation 2:} & \quad & a_{\rm fs} = 1 - 1/\kappa, \ a_{bs} = 0 ;
\eea
where $ \kappa \equiv \rho_0 J_z/2 K, \ \rho_0 = 1 / \pi u $.

{\it RKKY phase}: Let us start from Eq.(\ref{Model-1}) and derive the effective
Hamiltonian of KIs from the perturbation theory in $ J_\perp $. To this end,
we expand $ \exp(- \int {\rm d} \{x,\tau\} {\cal L}_{\rm bs}) $ up to $ O(J_\perp^2) $,
integrate over $ \phi $ and re-exponentiate the result. This yields the action
which describes spin interactions:
\be
\label{S_RKKY-nonloc}
  {\cal S} \!\! = \!\! -\frac{J_\perp^2}{(2 \pi \xi)^2} \! \sum_{j,j'} \int \!\! {\rm d} \tau_{1} {\rm d} \tau_{2} \
        S^+_j(\tau_1) \Pi(\tau_1 - \tau_2) S^-_{j'}(\tau_2);
\ee
$ \Pi $ is governed by the correlation function of the bulk bosons \cite{Giamarchi,NoFirstOrder}:
\be
\label{Pi}
   \Pi(t) =
     \left[
       \left( \frac{\beta u}{\pi \xi} \right)^2 \!\!
       \left( \!\!
         \sin^2( \pi t T) + \sinh^2 \!\! \left(\frac{x_j-x_{j'}}{L_T} \right) \!\!
       \right)
     \right]^{-\tilde{K}} \!\!\!\!\!\!\!\! .
\ee
Here $ \beta \equiv 1/T $; $ L_T \equiv \beta u / \pi $ is the thermal length. We will
consider the macroscopic spatial range $ \xi \ll R \ll L_T $. If
$ 1/2 < \tilde{K} \le 1 $, the main contribution to $ S_{j,j'} $ results from a small time
difference, $ T|\tau_1 - \tau_2| \sim |x_j-x_{j'}|/L_T \ll 1$.
This allows us to reduce $ {\cal S}_{j,j'} $ to the local action of RKKY:
\be
\label{S_RKKY-loc}
  {\cal S}_{\rm RKKY} = - E_{\rm RKKY} \int {\rm d}\, \tau
       \left[ S^+_1(\tau) S^-_2(\tau) + c.c. \right].
\ee
The terms with $ j = j' $ do not contribute to Eq.(\ref{S_RKKY-loc}) because $ S^+_j(\tau) S^-_j(\tau)
\propto ( \bar{d}_j + d_j )^2 = 0 $. We have introduced in Eq.(\ref{S_RKKY-loc}) the RKKY energy:
\be
\label{E_RKKY}
   E_{\rm RKKY} = \frac{2 J_\perp^2}{ (2 \pi \xi)^2 }
        \int_0^\beta {\rm d} t \ \Pi(t).
\ee
$ E_{\rm RKKY} $ can be expressed in terms of the hypergeometric functions. Its
asymptotic behavior for $ R / L_T \ll 1 $ is
\bea
\label{E_RKKY_asymp}
   E_{\rm RKKY} & \propto & \frac{J_\perp^2}{u \xi}
                  \!\!
                  \left[
     \frac{\Gamma\left(\frac{1}{2}-\tilde{K}\right)}{\Gamma(1-\tilde{K})} \left( \! \frac{\xi}{L_T} \! \right)^\alpha
         \!\!\! + \!
     \frac{\Gamma\left(\tilde{K}-\frac{1}{2}\right)}{\Gamma(\tilde{K})} \left( \! \frac{\xi}{R} \! \right)^\alpha
                  \right]; \cr
               \alpha & = & 2 \tilde{K} - 1.
\eea
If $ 1/2 < {\tilde K} < 1 $ and $ T \to 0 $, the first term in Eq.(\ref{E_RKKY_asymp}) vanishes
and the second one reproduces the usual RKKY energy. The failure of the paradigmatic theory
starts from $ \tilde{K} = 1/2 $ where both terms of Eq.(\ref{E_RKKY_asymp}) are needed to cancel
out divergences. Both contributions must be kept also at $ {\tilde K} < 1/2 $: neglecting the first
term leads to nonphysical results, like growth of $ E_{\rm RKKY} $ with increasing $ R $,
cf. Ref.\cite{RKKY-HLL-Int}. However, the first term diverges at $ \tilde{K} < 1/2 $ in the $ T
\to 0 $ limit. Moreover, the local time approximation used to derive Eq.(\ref{S_RKKY-loc})
loses its validity
because the UV singularity of $ \Pi $ becomes too weak and the integral is now given by all (not small)
time differences $ | \tau_1 - \tau_2 | < \beta $. All this signals that the physics changes at
the point $ \tilde{K} = 1/2 $ and the RKKY theory cannot be extended to smaller values of $ \tilde{K} $.

We emphasize the difference between Eq.(\ref{E_RKKY}) and its counterpart for
the spinful (almost) SU(2) symmetric Luttinger Liquid: in the latter case,
$ \Pi $ is a product of the charge and the spin sector contributions, $ \Pi_{c,s} $ with
the Luttinger parameters $ K_{c,s} $ \cite{RKKY-LL1}. This makes $ \Pi $ more singular at small times.
For example, if $ J_z = 0 $ and the electron interaction is SU(2) symmetric the exponent $ \tilde{K} $
in Eq.(\ref{Pi}) reduces to $ (K_c + 1)/2 > 1/2$, the integral in Eq.(\ref{E_RKKY}) converges at small times, the
theory is local in time and the effective Hamiltonian approach is valid.
Therefore, the above described crossover in the behavior of $ E_{\rm RKKY} $ is absent and a new
phase does not appear in the non-helical spinful Luttinger Liquid. Note that 1D interacting systems
driven far from the SU(2) symmetry may possess an emergent helicity with physics being
similar to that of our helical model \cite{Carr_2015}.

When the RKKY approach is valid
it is easy to calculate different spin correlation functions, e.g. $ G_{zz} =
- \hat{T}_\tau \la \hat{S}^z_1(\tau') \hat{S}^z_2(\tau) \ra, $ by using the effective
Hamiltonian, $ \hat{H}_{\rm RKKY} = - E_{\rm RKKY} ( \hat{S}^+_1 \hat{S}^-_2 + h.c. ) $,
which corresponds to the local action Eq.(\ref{S_RKKY-loc}). Calculations
at $ T \to 0 $ and the analytical continuation to the upper half-plane yield the
retarded Green's function:
\be
\label{Gzz_RKKY}
  G^R_{zz}(\omega) = - \frac{\pi}{2} \frac{|E_{\rm RKKY}|}{\omega_+^2 - (2 E_{\rm RKKY})^2}; \
  \omega_+ \equiv \omega + i 0.
\ee

{\it RKKY-Kondo transition}: To understand the transition to the new phase, let us switch
from the perturbation theory to the one-loop RG. We still work with the theory of Eq.(\ref{Model-1})
where the dimension of the backscattering vertex equals to $ \tilde{K} $. Thus, the leading
in $ J_\perp $ RG equation for this coupling constant reads as
\be
\label{RG_Jperp}
  \p_l J_\perp = (1 - \tilde{K}) J_\perp.
\ee
Here $ l $ is the logarithm of the energy $ \Omega $. The difference between RG for one \cite{MaciejkoLattice} and two
impurities is not visible at this level. Moreover, Eq.(\ref{RG_Jperp})
looks precisely like RG for the backscattering amplitude of the static impurities \cite{KaneFis:92a,KaneFis:92b},
though with renormalized $ K $. These two analogies are not accurate because the renormalization of
$ J_\perp $ stops quickly.

\begin{figure}[t]
\begin{center}
   \includegraphics[width=0.4 \textwidth]{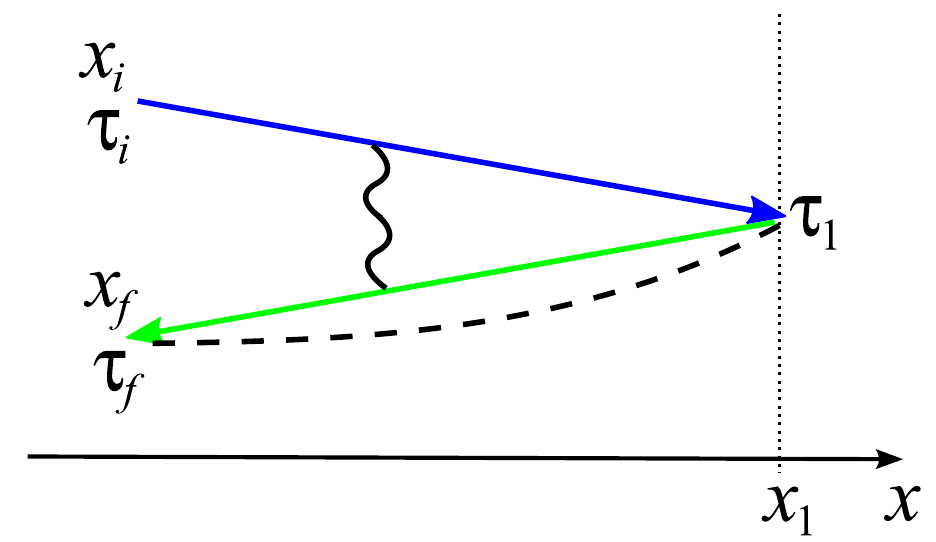}
\end{center}
\vspace{-0.25 cm}
   \caption{
        \label{Diagr}
        (color on-line) The diagram which yields the leading in interaction (wavy line) and in the
        coupling constants $ J_{\perp,z} $ correction to the Green function describing backscattering
        of a helical electron by a Kondo impurity. KI is located at the position $ x_1 $.
        Solid lines show the electron propagators before (blue) and after (green) backscattering. Dashed
        line denotes the spin propagator $ G_{-+} $ which stops logarithmic divergences of the
        theory of Refs.\cite{MatvGlazm_1993,MatvGlazm_1994} at $ E_{\rm RKKY }$.
           }
\end{figure}

Let us find the RG cutoff by adapting the scattering approach of Refs.\cite{MatvGlazm_1993,MatvGlazm_1994}
to the problem we study. The main idea of that approach is to consider the weak electron interaction and to find
logarithmic corrections to the Green's function of the backscattered electron, $ G_{\rm bs} = - \hat{T}_\tau
\la \hat{\psi}_L(\tau_f,x_f) \hat{\psi}^\dagger_R(\tau_i,x_i) \ra $. Here $ x_{i,f} \to - \infty $, $ \hat{\psi}^\dagger_R $
($ \hat{\psi}_L $) is the creation operator for right- (the annihilation operator for left-) moving fermion. The leading
correction to backscattering caused by the static impurity appears in the first order in the interaction,
$ \delta G_{\rm bs} \sim (1-K) \log(D/\Omega) $, $ \Omega > T $.

Now we recall that backscattering in HLL is caused by KI and requires spin-flip. Hence, the Green's function
describing backscattering must account for changing the spin state. Formally, one has to add the spin operator
in the definition of the Green's function:
\be
G^{\rm (KI)}_{\rm bs} = - \hat{T}_\tau \la \hat{S}^-(\tau_f) \hat{\psi}_L(\tau_f,x_f)
                                                                      \hat{\psi}^\dagger_R(\tau_i,x_i)
                                                                 \ra ;
\ee
the impurity number is omitted here.
The leading in $ (1 - K) $ and $ J_{\perp,z} $ correction to $ G^{\rm (KI)}_{\rm bs} $,
$ \delta G^{\rm (KI)}_{\rm bs} $, is given by the diagram shown in Fig.\ref{Diagr}. The difference
between $ \delta G_{\rm bs} $ and $ \delta G^{\rm (KI)}_{\rm bs} $ is due to the spin propagator
$ G_{-+} = - \hat{T}_\tau \la \hat{S}^-(\tau_f) \hat{S}^+(\tau_1) \ra $.
Using the parametrization Eq.(\ref{Diracs}), we obtain $ G_{-+} = - 2 \la c(\tau_f) \bar{c}(\tau_1) \ra
\la d(\tau_f) \bar{d}(\tau_1) \ra $; $ d $-fields have the bare Largangian
$ {\cal L}_f[\bar{d},d] $. Due to the inter-impurity correlations, the spin flip of one KI costs the energy of the gap
$ E_{\rm RKKY} $ which can be qualitatively described by adding the mass term to the Lagrangian of $ c $-fields:
$ {\cal L}_f[\bar{c},c] \to {\cal L}_f[\bar{c},c] + E_{\rm RKKY} \bar{c} c $. This yields:
\be
  G_{-+} = 2 \theta[ (\tau_f - \tau') E_{\rm RKKY}  ]  \, e^{ -  (\tau_f - \tau') E_{\rm RKKY} },
\ee
with the step function $ \theta(x \ge 0 ) = 1 $. $ G_{-+} $ changes the cutoff of the logarithm from $ \Omega $
to $ {\rm max}[ \Omega, E_{\rm RKKY} ] $.

The one-loop RG comes from re-summation of the leading logarithms. Therefore, we
conclude that RKKY correlations change the scale, at which the RG flow stops, from
a self-consistently obtained scale, $ E_{\rm sc} $, which marks the strong coupling
limit of the RG flow, to $ {\rm max}[ E_{\rm sc}, E_{\rm RKKY} ] $.
According to Eq.(\ref{RG_Jperp}), $ E_{\rm sc} $ coincides with $ T_{K} $
in the second line of Eq.(\ref{Scale_Tk}) with $ \tilde{K} $ being substituted for $ K $. The crossover
occurs at $ T_{K}(\tilde{K}) \sim E_{\rm RKKY} $ which obviously means the transition
between RKKY- and Kondo- physics at
\be
\label{CritLine}
  {\tilde{K}} = 1/2.
\ee
This explains failure of the paradigmatic theory for RKKY when $ {\tilde{K}} < 1/2 $.
The RKKY-Kondo transition is illustrated by the phase diagram in Fig.\ref{PhaseDiagr}.
We have restricted axes to the relevant range of $ K $ and $ | \rho_0 J_z | < 1 $ and
have excluded the extremely strong coupling and the second critical line from
this figure. The phase diagram of Fig.\ref{PhaseDiagr} is different from that for the
single KI \cite{MaciejkoLattice}: the border between two phases is defined by
Eq.(\ref{CritLine}) for two KIs while by $ \tilde{K} = 1 $ for the single KI.

{\it Kondo phase}: Two impurities coupled to HLL is not the exactly solvable model
and, therefore, one cannot say much about the Kondo phase without numerics.
One possibility for analytics is provided by a vicinity of the so-called decoupling
limit \cite{MaciejkoLattice} which can be conveniently analyzed by using
Eq.(\ref{Model-2}) with $ |a_{\rm fs}| \ll 1 $ \cite{TwoModels}. In this case, the spin Green's
function $ G_{zz} $ can be calculated perturbatively in $ (\rho_0 J_z a_{\rm fs}) $ and
exactly in $ J_\perp $. Similar to Eq.(\ref{Gzz_RKKY}), we do the analytic
continuation to the upper half-plane at $ T \to 0 $ and find $ G^R_{zz} $ near the
decoupling limit:
\be
\label{Gzz_DL}
  G^R_{zz}(\omega) \simeq i \left( \frac{\pi}{2} \right)^3 \left( \rho_0 J_z a_{\rm fs} \right)^2 \!\!
              \left( \frac{\Omega_\perp}{\omega_+^2 - \Omega_\perp^2} \right)^2 \!\!
              \frac{\omega}{K} e^{i \frac{R \omega_+}{u}};
\ee
with $ \Omega_\perp \equiv J_\perp / 2 \pi \xi $.

The difference between two phases becomes obvious after comparing the frequency dependence
of $ G^R_{zz} $ in Eqs.(\ref{Gzz_RKKY}) and (\ref{Gzz_DL}). In the Hamiltonian description
of the RKKY phase, there is no retardation and $ G^R_{zz} $ becomes constant at
$ |\omega| \ll E_{\rm RKKY} $. This reflects the RKKY-induced inter-impurity correlation.
The retardation is present in Eq.(\ref{Gzz_DL}) (note the oscillating exponential) and,
much more importantly, $ G^R_{zz} $ decays as $ \omega / \Omega_\perp $ at $ |\omega| \ll
\Omega_\perp $. This decay shows the absence of the noticeable inter-impurity correlation
near the decoupling limit. When $ \omega \to 0 $, i.e., the observation time goes to
infinity, (almost) uncorrelated dynamics of two KIs leads to the the suppression of $ G^R_{zz} $.
If the inter-impurity correlation is weak we can make use of the RG for the single KI which
shows the flow toward the decoupling limit where KIs are not correlated and only the
Kondo-like backscattering remains relevant \cite{MaciejkoLattice}.

All these observations confirm that the Kondo physics fully dominates at $ \tilde{K} < 1/2 $.

{\it To summarize}, we have shown that the paradigmatic RKKY theory is not
applicable if the indirect exchange interaction of two spin-1/2 Kondo impurities
is mediated by strongly correlated helical electrons with the effective Luttinger parameter
$ \tilde{K} < 1/2 $, Eq.(\ref{K_Eff}). The physical reason for
this counterintuitive finding is the competition between RKKY induced spin
correlations and Kondo screening of localized spins. This competition is crucially 
intensified by helicity.
Phenomenological arguments combined with the perturbation theory
and with a scaling analysis of the one-loop Renormalization Group have allowed
us to identify a border between phases where either the RKKY- or the
Kondo physics dominates, Fig.\ref{PhaseDiagr}. These phases emerge
when the (effective) electron interaction is weak or strong, respectively.

We have encountered an instructive example of the interacting system where the usual
description of a subsystem in terms of an effective Hamiltonian is impossible due to
helicity and strong interaction. Physical situations where the effective Hamiltonian of a
subsystem cannot be constructed put forward a conceptual problem of treating such
strongly correlated systems.

Our results give a new insight into the fundamental phenomenon of the RKKY-Kondo competition.
In particular, they indicate that the Doniach phase diagram \cite{Doniach}
can be very non-trivial in systems with spin-orbit interaction.
This famous diagram describes a crossover of a Kondo lattice between magnetically
ordered phases and phases of heavy fermion Fermi liquid which are dominated by
correlations between local magnetic moments and by the Kondo screening, respectively.

Our predictions may serve as a basis for describing an influence of a rare 
Kondo array on transport in helical systems.
Measurements of Ref.\cite{Li_2015} suggest that HLL on the edges of 2D topological insulators
made of InAs/GaSb can have really small Luttinger parameter, $ K \sim 0.2 < 1/2 $. We thus expect
that our predictions are relevant for the experimental studies of the topological
insulators. Another possible platform, where the unusual RKKY-Kondo competition can be detected,
is provided by recently fabricated 1D wires with interactions induced helicity
\cite{Quay_2010,Scheller_2014,Heedt_2017,Kammhuber_2017}.
Further development of the theory may include a detailed study of
a vicinity of the transition and an extension to the case of larger spins.

\begin{acknowledgments}
{\bf Acknowledgments}:
We gratefully acknowledge hospitality of the Abdus Salam ICTP and of  the IBS Center for Theoretical
Physics of Complex Systems (Daejeon, South Korea) where the part of this project was done.
V.I.Yu. acknowledges the partial support of RFBR grant 15-02-05657 and the Basic Research
Program of HSE. O.M.Ye. acknowledges the support from the DFG through the grant YE 157/2-1.
We are grateful to Boris Altshuler, Jan von Delft and Ari Wugalter for useful discussions.
\end{acknowledgments}

\bibliography{Bibliography,2ImpKondo,Spin-Params,FootNotes,AddRefs}

\begin{thebibliography}{61}
\bibitem [{K-r()}]{K-range}%
  \BibitemOpen
  \href@noop {} {}\bibinfo {note} {We restrict ourselves to the moderately
  strong repulsion. If the electron repulsion is extremely strong, $ K < 1/4 $,
  one has to take into account multi-particle scattering processes
  \cite{MaciejkoOregZhang} which are beyond our consideration}\BibitemShut
  {NoStop}%
\bibitem [{com()}]{comm-K}%
  \BibitemOpen
  \href@noop {} {}\bibinfo {note} {The transition was noticed in
  Ref.\cite{RKKY-HLL-Int} though it was neither properly treated nor correctly
  explained since the authors of that paper had erroneously extended the fully
  perturbative theory beyond its validity range to $ 1/4 < K \le 1/2 $. This
  resulted in nonphysical answers, like growth of $ E_{\rm RKKY} $ with
  increasing $ R $.}\BibitemShut {Stop}%
\bibitem [{Sz-()}]{Sz-const}%
  \BibitemOpen
  \href@noop {} {}\bibinfo {note} {We note in passing that the constant 1/2 in
  the expression for $ S^z $ is important in the operator language but it
  yields only unimportant boundary term in the path integral formulation of our
  model.}\BibitemShut {Stop}%
\bibitem [{Spi()}]{Spins-RKKY}%
  \BibitemOpen
  \href@noop {} {}\bibinfo {note} {Simpler representations, where the
  redundancy of the fermionic fields is reduced (e.g. $ S_{1,2}^+ = [ \bar{a}
  \pm a ] \bar{c}_{1,2} $), do not reproduce the RKKY action
  correctly.}\BibitemShut {Stop}%
\bibitem [{NoF()}]{NoFirstOrder}%
  \BibitemOpen
  \href@noop {} {}\bibinfo {note} {First order terms do not appear in the
  thermodynamic limit because of the so-called ``electroneutrality property''
  \cite{Giamarchi}.}\BibitemShut {Stop}%
\bibitem [{Two()}]{TwoModels}%
  \BibitemOpen
  \href@noop {} {}\bibinfo {note} {Whereas the 1st representation,
  Eq.(\ref{Model-1}), has allowed us to obtain the RKKY action rather simply
  the 2nd one, Eq.(\ref{Model-2}), is not suited for this procedure because,
  after integrating out the bulk bosons, one has to average the perturbative
  expansion in $ J_\perp $ over the action with the $ S_1^z $-$ S_2^z $
  interaction.}\BibitemShut {Stop}%
\end{thebibliography}


\end{document}